\documentclass{aa}

\def\lsim{\mathrel{\rlap{\lower4pt\hbox{\hskip1pt$\sim$}}
    \raise1pt\hbox{$<$}}}         
\def\gsim{\mathrel{\rlap{\lower4pt\hbox{\hskip1pt$\sim$}}
    \raise1pt\hbox{$>$}}}         
\def\esim{\mathrel{\rlap{\raise2pt\hbox{$\sim$}}
    \lower1pt\hbox{$-$}}}         

\usepackage{graphics}

\begin{document}

\thesaurus{09.03.3; 12.04.1; 10.08.1}

\title{Signatures of exotic physics in antiproton cosmic ray 
measurements}

\author{Piero Ullio}
\institute{Department of Physics, Stockholm University, Box 6730,
SE-113~85~Stockholm, Sweden
\thanks{E-mail: piero@physto.se}}

\date{Received / Accepted }

\maketitle

\begin{abstract}
   
More than a decade ago it was noticed that  
an unexpectedly large value of the measured cosmic antiproton
flux at low kinetic energies could be interpreted in terms of
a neutralino-induced component.
An overproduction of low energy antiprotons seems however 
to be disfavoured by recent data from the {\sc Bess} experiment. 
The strategy we propose here is to focus instead on the high 
energy antiproton spectrum. We find cases in which the signal 
from neutralino annihilations in a clumpy halo scenario can be 
unambiguously distinguished from the cosmic ray induced component.
Such signatures can be tested in the near future by upcoming 
experiments. 
We investigate also the possibility that antiprotons have 
a finite lifetime. We find cases in which the neutralino-induced flux 
is consistent with present data for $\tau_{\bar{p}}$ as low as 
0.15~Myr, in clear violation of a recent bound derived from 
the antiproton cosmic ray flux.

\end{abstract}


\section{Introduction}

The idea of exploiting cosmic antiprotons measurements to probe
unconventional particle physics and astrophysics scenarios is
certainly not new. 
The investigation on exotic antiproton sources
was stimulated by early reports (Golden et al. 1979; 
Buffington et al. 1981) of unexpectedly large values for the 
antiproton to proton ratio in cosmic rays.

A possibility which has been discussed in some details is that a 
significant contribution to the antiproton flux may be given by 
pair annihilation of relic neutralinos. The lightest neutralino,
plausible the lightest supersymmetric particle in the Minimal
Supersymmetric extension to the Standard Model (MSSM), 
is one of the leading dark matter candidates. 
Its coupling with ordinary particles has the strength of
weak interaction.
This is exactly what it is needed to provide a relic density of
the right order for a flat Universe.
It implies as well that, if one invokes a population of relic 
neutralinos in the halo of galaxies to solve the dark matter problem,
these neutralinos can annihilate in pairs into standard model
particles, which might eventually be detected. 
In particular it has been shown that the neutralino-induced 
antiproton flux can be
at the level of the background flux from cosmic rays
(see e.g. Silk \& Srednicki (1984); Stecker et al. (1985); 
Bottino et al. (1998); Bergstr\"om et al. (1999b)).
The question to address is of course if there are distinctive
features in this exotic component so that it can be singled
out unambiguously.

The ordinary cosmic ray induced antiprotons are generated in
collisions of primary particles with the interstellar medium.
The main source are $pp$ collisions which, for kinematical reasons,
produce a characteristic spectrum peaked at a kinetic energy
of about 2~GeV and rapidly decreasing at low energies.
The component from neutralino annihilations may not fall as fast; 
in some cases its maximum is actually in the low energy region. 
This is the signature which was proposed more than a decade ago
in connection with early antiproton measurements
(Silk \& Srednicki 1984; Stecker et al. 1985). 
The more recent (and probably more accurate) data from the {\sc Bess} 
experiment (Matsunaga et al.\ 1998; Orito 1998) indicate 
unfortunately that the low energy flux may not be so high as 
previously thought. It actually seems to be at a level that
is in very good agreement with a standard prediction for the 
secondary flux if one takes into account $pHe$ collisions and 
energy losses during antiproton propagation  (see 
Bergstr\"om et al. 1999b, hereafter Paper I). If future 
measurements will confirm {\sc Bess} data, this will lead inevitably to
an impasse in this neutralino detection technique: 
if the measured flux is consistent with some estimate of the 
background, there is no way in improving the signal to background 
ratio.

There are however signatures of a neutralino-induced flux
which are alternative to the one we have just discussed. The strategy 
we propose here is to search for an exotic component
in the high energy antiproton spectrum, 
above the expected maximum of the background.
This kind of investigation is motivated by the fact that
the secondary antiproton flux at high energies is predicted
with great confidence.
It is remarkable that in this energy region nearly all estimates
in the literature are consistent with each other. 
The three main ingredients to compute this part of the antiproton
spectrum, unlike in the low energy range, are actually very well 
established: data on scattering of protons on nuclei are sufficiently
aboundant, the spectral index of the primary proton flux is known 
with good accuracy and, finally, cosmic ray measurements fix quite 
strictly the 0.6 power law scaling of the diffusion coefficient 
with rigidity. It follows then that at least the shape, if not the
normalization, of the interstellar antiproton flux at high kinetic 
energies is significantly constrained. 
Moreover in this regime solar modulation, which introduces a further 
factor of uncertainty at low energies, does not play a main role.

It was shown in Paper~I that in the canonical scenario
in which neutralinos are assumed to form a perfectly smooth ``gas''
of dark matter particles extended throughout the halo of the Galaxy, 
the induced antiproton flux is not sufficiently enhanced to cause 
significant distorsions to the high energy secondary spectrum. 
As we will show in the next Section a scenario with a clumpy halo 
can be much more favourable. 

We have singled out three classes of examples in which 
the signal we propose can be unambiguously distinguished from the 
secondary component. In the first two the main ingredient is clumpy
neutralino dark matter, in the third we consider also the exotic
possibility that antiprotons have a finite lifetime. 
Before describing these three cases, we introduce the formalism
needed to give predictions for antiproton fluxes from clumps of
neutralino dark matter in the Milky Way.

\section{Neutralino-induced antiproton flux in a clumpy halo scenario}

\subsection{General discussion}

There are reasons to question whether the distribution of dark matter
in galactic halos has to be regarded as smooth on all scales.
Several theoretical models predict that dark matter may
aggregate in high density substructures, ``clumps'' of dark matter
(see e.g. Silk \& Szalay (1987); Silk \& Stebbins (1993);
Kolb \& Tkachev (1994)).
From the phenomenological point of view, as the annihilation rate 
depends quadratically on the neutralino number density locally in 
space, the presence of such clumps in the Galaxy may significantly 
enhance the chances of detecting neutralino dark matter.
If we postulate that at least a fraction of the dark matter in the 
Milky Way is clustered in high density regions, we can associate 
to a given clump of density profile
$\rho_{cl}(\vec{r}_{cl})$ and located at the position $\vec{x}_{cl}$,
the following antiproton source term:
\begin{eqnarray}
  Q_{\bar{p}}^{\chi,\,cl}(T,\vec{x}\,) &=&
  (\sigma_{\rm ann}v)
  \frac{1}{{m_{\chi}}^{2}} \sum_{F}^{}\frac{dN^{F}}{dT}B^{F} \cdot 
\nonumber \\
&& \cdot \int d^{\,3}r_{cl}\,\left(\rho_{cl}(\vec{r}_{cl})\right)^{2}
  \; \delta^3\left(\vec{x}-\vec{x}_{cl}\right)\;.
  \label{eq:sourcl}
\end{eqnarray}
The first terms on the right hand side of this equation are related 
to the particle physics dark matter candidate. 
In our case, $\sigma_{\rm ann}v$ is the total annihilation rate of 
non relativistic neutralinos, while 
$m_{\chi}$ is the neutralino mass (which enters in the equation 
with a -2 power as the dark matter density divided by $m_{\chi}$
is the neutralino number density). We then sum over the annihilation
final states $F$ which can give antiprotons in a fragmentation or 
decay process, namely heavy quarks, gluons, gauge and Higgs bosons.
For each of them, $B^{F}$ and $dN^{F}/dT$ are, respectively,
the branching ratio and the fragmentation function.
The hadronization of all final states has been simulated 
with the Lund Monte Carlo program {\sc Pythia} 6.115 
(Sj\"ostrand 1994).

$\sigma_{\rm ann}v$ and $B^{F}$ are computed in the framework of the
MSSM as described in Paper~I.
Both the value of the annihilation cross section and the relative 
importance of different final states can vary largely in different 
regions of the MSSM parameter space.
The phenomenological version of the MSSM we use is described in
Bergstr\"om \& Gondolo (1996).
We just remind here few definitions we need in the discussion in 
Section~\ref{sec:high}. The lightest neutralino is the lightest
mass eigenstate obtained from the superposition of the 
superpartners $\tilde{B}$ and $\tilde{W}^3$ of the neutral 
gauge bosons, 
and of the neutral CP-even Higgsinos $\tilde{H}^0_1$ and 
$\tilde{H}^0_2$.
The neutralino is usually defined
Higgsino-like if it is mainly a linear combination of 
$\tilde{H}^0_1$ and $\tilde{H}^0_2$, 
while it is called gaugino-like if it is mainly given by the
superposition
of the $\tilde{B}$ and $\tilde{W}^3$ interaction eigenstates.
The formal classification is done introducing the gaugino 
fraction $Z_g$.

The remaining terms in the expression for the antiproton source 
function, Eq.~(\ref{eq:sourcl}), are related to the neutralino
density distribution; we have actually assumed that the extension of 
clumps is much smaller than galactic scales. 
The source is hence treated as pointlike: this is not strictly 
necessary but simplifies the discussion which follows. 

The propagation of antiprotons in the Galaxy is treated with a
two zone diffusion model. Assuming a cylindrical symmetry for the 
Galaxy, a solution to a transport equation of the diffusion type
can be derived introducing a Bessel-Fourier series to factorize out 
the radial dependence and then solving a one dimensional differential 
equation in the vertical direction.
Details on the model and an analytic 
solution in case of a pointlike source are given as well in Paper~I.
As the diffusion equation we apply is linear in the antiproton 
number density $N$, we can consider each source separately and then 
sum the contributions to compute $N$.

To characterize clumps, it is useful to introduce the dimensionless 
parameter
\begin{equation}
    \delta =\frac{1}{\rho_{0}} 
            \frac{\int d^{\,3}r_{cl}\,
                  \left(\rho_{cl}(\vec{r}_{cl})\right)^{2}}
                  {\int d^{\,3}r_{cl}\,\rho_{cl}(\vec{r}_{cl})}
\end{equation}
which gives the effective contrast between the dark matter density 
inside the clump and the local halo density $\rho_0$. We introduce 
this definition in Eq.~(\ref{eq:sourcl}) and then, switching to 
cylindrical coordinates, we derive the coefficients of the 
Bessel-Fourier series for the source function
(see Eq.~(18) in Paper~I). We find:
\begin{eqnarray}
  \lefteqn{Q_s^k(z) = (\sigma_{\rm ann}v)
  \frac{1}{{m_{\chi}}^2} \sum_{F}^{}\frac{dN^{F}}{dT}B^{F}
  \; \rho_0\,M_{cl}\,\delta\,\frac{1}{\alpha_k\,\pi}\cdot} 
  \nonumber \\
&& 
  \cdot \frac{2}{{R_h}^2\,{J_{k+1}}^2(\nu_s^k)}
  \,J_k \left(\nu_s^k \frac{r_{cl}}{R_h}\right)
  \,\cos(k\,\theta_{cl})\,\delta(z-z_{\rm{cl}})
\label{eq:clumpsou}
\end{eqnarray}
where $M_{cl}$ is the mass of the dark matter source we are 
considering.
This formula has to be substituted into Eq.~(21) and 
(23) of Paper~I to derive the antiproton number density and flux.
It is useful to factorize out in the expression for the flux the 
dependence on the MSSM parameters. We define, in analogy to the
expression for a smooth halo scenario, 
\begin{equation}
  \Phi_{\bar{p}}^{cl}(r_0,T) \equiv
  (\sigma_{\rm ann}v) \sum_{F}^{}\frac{dN^{F}}{dT}B^{F}
  \frac{\rho_0\,M_{cl}\,\delta}{{m_{\tilde{\chi}}}^{2}}
  \, C_{\rm prop}^{cl}(T,\vec{x}_{cl})
  \label{eq:signalcl}
\end{equation}
where the coefficient $C_{\rm prop}^{cl}$, which has the dimension
of a length to the power $-2$ times a solid angle to the power $-1$,
contains the dependence of the flux on the neutralino distribution
locally in space and on the parameters in the propagation model. 
There is a one to one correspondence between the 
antiproton flux computed summing the contributions from the smooth
halo scenario as done in Paper~I, and that from a single clump,
Eq.~\ref{eq:signalcl}; one has just to
replace the coefficient $C_{\rm prop}$ in Eq.~(40) of Paper~I
with $C_{\rm prop}^{cl} \cdot M_{cl}\,\delta / \rho_0$.
To facilitate the comparison we will plot $C_{\rm prop}^{cl}$ in 
units of $10^{24}\,\rm{cm}\,\rm{sr}^{-1}\,\rm{kpc}^{-3}$.

\subsection{Antiproton flux from single clumps}

\begin{figure}[tb]
 \resizebox{\hsize}{!}{\includegraphics{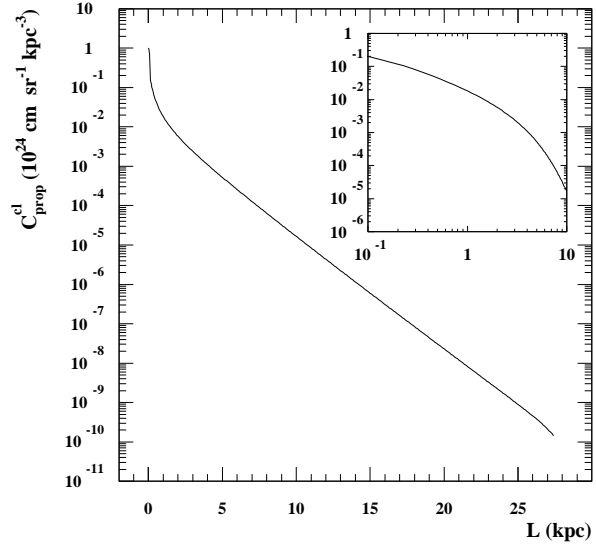}}
\caption{$C_{\rm prop}^{cl}$ at T=1~GeV and as a function of the distance
from the observer $L$.}
\label{fig:cprpopcl}
\end{figure}

We first suppose that there are few heavy clumps and study their 
individual influence on the result for the antiproton flux. Given 
the factorization introduced in Eq.~(\ref{eq:signalcl}), we just 
analyse the coefficient $C_{\rm prop}^{cl}$ and its behaviour as 
a function of the position of the source in the Galaxy. Intuitively we
expect that sources which are nearby give the largest contributions to 
the flux. In case of a propagation model which can be regarded to 
be as for an infinitely large diffusion box, we would find that 
$C_{\rm prop}^{cl}$, rather than being a function of the three 
coordinates $\vec{x}_{cl}$, depends just on the distance from the 
observer, the line of sight distance $L$. The boundary conditions
in the propagation model, i.e. free escape at border of the diffusion
region, break strictly speaking this symmetry; their effect is 
however not very severe for sources which are not too close to the 
border of the diffusion zone (in the vertical direction not closer 
than 0.8--1~kpc).

\begin{figure}[tb]
 \vspace{1 cm}
 \centerline{\resizebox{0.8\hsize}{!}{\includegraphics{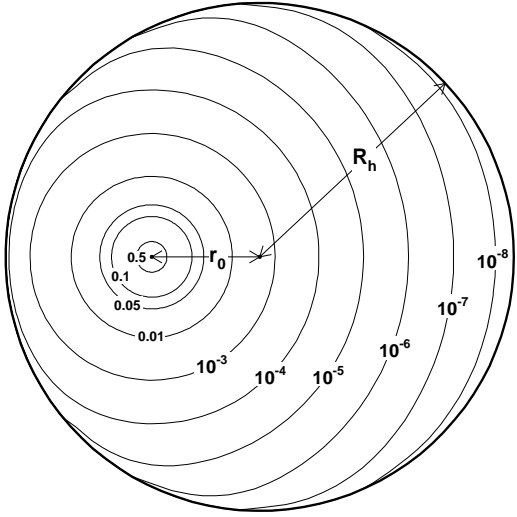}}}
 \vspace{0.75 cm}
 \caption{Isolevel curves for $C_{\rm prop}^{cl}$ in the 
  plane $z_{cl}=1$~kpc. $C_{\rm prop}^{cl}$ has been normalized to~1 
  at 1~kpc above our position in the Galaxy. $r_0$ is our 
  galactocentric distance.}
\label{fig:level}
\end{figure}

In Fig.~\ref{fig:cprpopcl} we plot $C_{\rm prop}^{cl}$ as a function 
of the distance $L$, choosing T=1~GeV and the set of canonical
parameters for propagation model introduced in Paper~I (diffusion zone
height $h_h = 3 \,\rm{kpc}$, constant term in the diffusion coefficient
$D^0 = 6 \cdot 10^{27}\,\rm{cm}^2 \rm{s}^{-1}$ and rigidity scale
in the diffusion coefficient
$R_0 = 3 \,\rm{GV}$). The function we have drawn was computed 
fixing $z_{cl}$ and moving away from our position in the Galaxy in 
the direction of the galactic 
centre.\footnote{Details on how to implement the 
numerical formula to compute $C_{\rm prop}^{cl}$ are given in
Ullio (1999).}
As can be seen, we find at large distances a very accurate 
exponential scaling, while a sharp cusp appears below a distance of 
few kiloparsecs: the intuition that if the source happens to be 
close to us it gives a large contribution is confirmed. 
To show how accurate the assumption is that $C_{\rm prop}^{cl}$ is 
essentially just a function of $L$, we plot in Fig.~\ref{fig:level}
its isolevel curves in the plane $z_{cl}=1$~kpc. 
We have chosen to normalize $C_{\rm prop}^{cl}$ to 1 at 
1~kpc above our position in the Galaxy. 
The isolevel curves are basically circumferences, which are just 
slightly deformed by our choice of fixing the antiproton number 
density to be zero at the border of the diffusion zone in the radial 
direction, i.e. $r=20$~kpc (thick solid line in the figure).

We will not examine here the dependence of $C_{\rm prop}^{cl}$, and
hence of the flux, on the parameters which define the propagation 
model: the results are analogous to what it was derived for 
$C_{\rm prop}$ in the smooth halo case. We just mention that 
introducing a galactic wind fixes a preferred direction of 
propagation and breaks therefore the scaling with distance.

As an example we consider the contribution to the antiproton flux 
given by a clump of neutralino dark matter defined by the same 
choice of parameters as in Bengtsson et al.~(1990), where clumpiness 
was introduced in connection with the neutralino induced 
$\gamma$-ray signal with continuum energy spectrum.
We fix hence $M_{cl} \sim 10^{8} M_\odot$ and $\delta \sim 10^3$;
the prefactor $M_{cl} \delta$ is then about 
$1.3 \cdot 10^4 \rho_0\,\rm{kpc}^{3}$, where we assumed as local 
halo density $\rho_0 = 0.3 \,\rm{GeV}\,\rm{cm}^{-3}$. 
Comparing the coefficient $C_{\rm prop}^{cl}$ with the analogous 
quantity in a smooth halo scenario (with the same choice
of propagation model parameters 
$C_{\rm prop}(T=1\,\rm{GeV}) \sim 10^{25}\,\rm{cm}\,\rm{sr}^{-1}$)
we find that the antiproton flux from this single dark matter clump 
can be at the level (or much higher) than the sum of the 
contributions from the whole dark matter halo if this source is 
within about~4.5~kpc.

We might also consider an opposite approach. It is well established 
that a dark mass of at least $2 \cdot 10^{6} M_\odot$ is concentrated 
within 0.015~pc at the galactic centre (Eckart \& Genzel 1996), 
forming probably a massive black hole, possibly the astrophysical 
object which is called Sgr A$^*$. Such accretion of matter might be 
associated to a region where the density of neutralino dark matter 
is enhanced as well.
In an extreme scenario (Berezinsky et al. 1992) the potential well of 
a very steep dark matter halo profile ($\sim 1/r^{1.8}$) is the seed 
for the formation of the black hole itself. It is also intriguing
that an excess in the high energy $\gamma$-ray flux from the 
galactic centre region, which has been found from the analysis of
EGRET data (Mayer-Hasselwander et al. 1998), 
can be explained in terms of neutralino annihilations if an appropriate
enhancement of the neutralino density is present there (Ullio 1999).
Regardless of its possible origin, we can estimate how large the 
accretion of neutralinos at the galactic centre should be to give a 
measurable primary antiproton flux.
Assuming that the galactocentric distance is 8.5~kpc, we find that the 
flux induced by a source at the galactic centre is proportional to
$C_{\rm prop}^{cl} (8.5\,\rm{kpc}) \sim 4.5 \cdot 10^{19}
\,\rm{cm}\,\rm{sr}^{-1}\,\rm{kpc}^{-3}$ (at T=1~GeV ).
If we require for instance that its contribution should be at least 
one half of the total flux in a smooth halo scenario, we find that   
$M_{cl} \delta$ should be at least of the order of $10^{12} M_\odot$.

\subsection{Collective effects of clumpiness}

There is the possibility that a large fraction of the dark matter 
mass is in clumps, in the extreme case all of it. 
To avoid violating dynamical constraints,
clumps should be light, with masses probably less than 
$M_{cl} \sim 10^{4} - 10^{6}  M_\odot$.
If one could deal with a model that gives some accurate prediction 
for the masses of clumps and their distribution in the Galaxy, it 
would be possible to exploit the approach of the previous paragraph 
and estimate the antiproton flux by adding the contributions from 
individual sources.
As very little is known about the inherently non linear problem of
generating dark matter clumps, it seems more reasonable to follow a 
probabilistic approach.

Let $f$ be the fraction of dark matter in clumps and $N_{cl}$ the total
number of clumps, all roughly of about the same mass and overdensity.
We can define a probability density distribution of the clumps in the 
Galaxy which in the limit of large $f$, to fulfill 
dynamical constraints, has to follow the mass distribution in the 
halo. In a Cartesian coordinate system with origin at the galactic 
centre, the probability to find  a given clump in the volume 
element $d^{\,3}x$ at position $\vec{x}$ is:
\begin{equation} 
  p_{cl}(\vec{x})\,d^{\,3}x = \frac{1}{M_h}\; \rho(\vec{x})\,
  d^{\,3}x\;.
\label{eq:prob} 
\end{equation} 
We introduced here $M_h$, the total mass of the halo, so that $p_{cl}$
has the correct normalization $\int p_{cl} (\vec{x}) d^{\,3}x = 1$.

The antiproton source function in the volume element $d^{\,3}x$ at the 
galactic position $\vec{x}$ is then:
\begin{eqnarray}
  \lefteqn{Q_{\bar{p}}^{\chi\,\prime}(T,\vec{x}\,)\,d^{\,3}x = 
  (\sigma_{\rm ann}v)
  \frac{1}{{m_{\chi}}^{2}} \sum_{F}^{}\frac{dN^{F}}{dT}B^{F} \cdot}
  \nonumber \\
&& 
  \cdot N_{cl}\,p_{cl}(\vec{x}) \int d^{\,3}r_{cl}\,
  \left(\rho_{cl}(\vec{r}_{cl})\right)^{2} \, d^{\,3}x
  \nonumber \\
&&
  = f\,\delta\;\frac{\rho_0\;\rho(\vec{x})}{{m_{\chi}}^{2}} 
  (\sigma_{\rm ann}v) \sum_{F}^{}\frac{dN^{F}}{dT}B^{F} \, d^{\,3}x \;,
  \label{eq:srclhalo}
\end{eqnarray}
while the antiproton flux is:
\begin{eqnarray}
  \lefteqn{\Phi_{\bar{p}}^{\prime}(r_0,T) = 
  (\sigma_{\rm ann}v) \sum_{F}^{}\frac{dN^{F}}{dT}B^{F}
  f\,\delta\; \frac{{\rho_0}^2}{{m_{\tilde{\chi}}}^{2}}\cdot}
  \nonumber \\
&&
  \cdot\,\int d^{\,3}x\, \frac{\rho(\vec{x})}{\rho_0}
  C_{\rm prop}^{cl}(T,\vec{x})
  \nonumber \\
&&
  \equiv   (\sigma_{\rm ann}v) \sum_{F}^{}\frac{dN^{F}}{dT}B^{F}
  f\,\delta\; \frac{{\rho_0}^2}{{m_{\tilde{\chi}}}^{2}}
  C_{\rm prop}^{\prime}(T) \;.
  \label{eq:fluxcl}
\end{eqnarray}
In the equation above, $C_{\rm prop}^{\prime}(T)$ may actually be 
computed in the same way as the corresponding coefficient for
a smooth halo $C_{\rm prop}(T)$, just by replacing in the 
former $(\rho(\vec{x})/\rho_0)^2$ with $\rho(\vec{x})/\rho_0$.
The two coefficients have similar behaviours; for our canonical 
diffusion parameters and an isothermal sphere as dark matter halo
profile, $C_{\rm prop}^{\prime} \sim 0.75 \cdot C_{\rm prop}$ 
for most of the kinetic energies of interest in our problem. 
The antiproton flux in the many small clump scenario can then be 
obtained by scaling the result in the smooth halo case by roughly 
$0.75 \cdot f\,\delta$. There is quite a large freedom in the 
choice of $f\,\delta$, however one has to worry about not violating 
existing experimental bounds (Bergstr\"om et al. 1999a). 
In all the results in the next Section we have been careful to check 
that the models we propose do not induce an overproduction of both 
diffuse $\gamma$-rays and of cosmic positrons.

\section{Signatures in the high energy antiproton spectrum}
\label{sec:high}

We have provided two schemes in which the antiproton flux can be 
sensibly enhanced with respect to the results in the smooth halo 
scenario. We propose here 
three cases in which the signal from neutralino dark matter 
annihilations  in a clumpy halo can distort the high 
energy cosmic ray flux, giving a very clear signature
of the presence of an exotic component.

Two experiments have measured the antiproton flux above a 
kinetic energy of few GeV (Golden et al. 1979; Hof et al. 1996) 
and their data are in contradiction with each other. A new set of 
experiments, the {\sc Caprice} experiment (Boezio et al. 1997)
and the space-based {\sc Ams} (Ahlen et al. 1994)
and {\sc Pamela} (Adriani et al. 1995),
should give much more abundant data in the near future.
We will check that our predictions are consistent with the data in 
the low energy regime (below 3~GeV) from the {\sc Bess}~97 
(Orito 1998) and {\sc Bess}~95 (Matsunaga et al.\ 1998)
flights, which have given a first hint on the actual shape of the 
antiproton spectrum. The prediction for the background for our 
canonical set of parameters actually provides a very good fit of 
the data (see Paper~I; our background prediction has 
been confirmed by Bieber et al. (1999) and is consistent
in the high energy region with most the previous estimates in 
the literature).
In the first two cases below we allow a little room 
for a neutralino-induced component by lowering the normalization
of the primary proton flux by 1~$\sigma$. 

For simplicity we will focus on the many small clumps scenario 
and quote in each case the value of the parameter $f\delta$ we are 
considering. 
The sample of supersymmetric models on which this analysis is based
is the same as in Paper~I. As in the smooth halo case we
restrict to those MSSMs for which the neutralino has a cosmologically
interesting relic density, e.g. $0.025 < \Omega_{\chi}h^{2} < 1$. 

To compare with {\sc Bess} data we have to take into account
solar modulation. We relate the flux measured at the top of the
atmosphere to the interstellar flux applying the force-field
approximation by Gleeson \& Axford (1967) with solar modulation
parameter $\phi_F \sim 500$~MV (value suggested in the analysis
of the {\sc Bess} collaboration).

\subsection{Broadening of the maximum in antiproton flux}

\begin{figure}[tb]
\resizebox{\hsize}{!}{\includegraphics{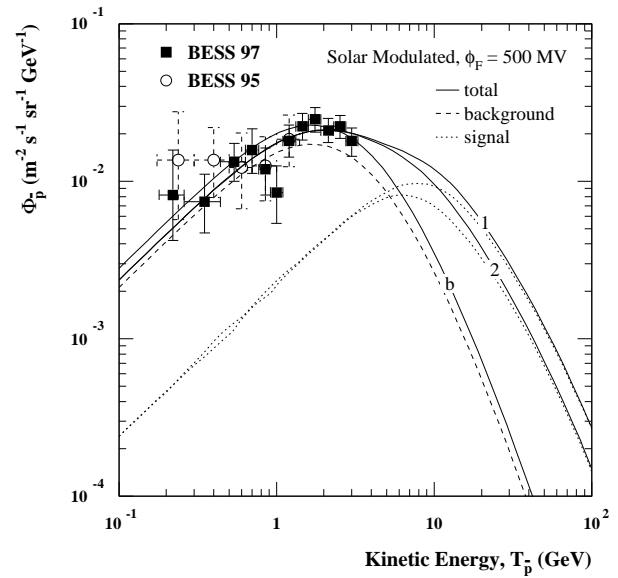}}
\caption{Distorsion on the secondary antiproton flux (line $b$)
induced by a signal from a heavy Higgsino-like neutralino.}
\label{fig:speccl1}
\end{figure}

The prediction for the background secondary flux has a well defined 
maximum at about 2~GeV. We check if it is possible to introduce a 
distorsion in the spectrum by adding a component which widens the 
shape of the maximum. 
We will further require that this is not just some local distorsion 
which might be mimicked by introducing for instance 
reacceleration effects (Simon \& Heinbach 1996). 
In our extreme case we impose that the antiproton flux from 
neutralino annihilations must exceed the estimate for the background 
by at least one order of magnitude at very high energies,
say about 50~GeV. To obtain a wider maximum one has to add a 
neutralino signal rather sharply peaked at an energy higher than the 
maximum in the background; its spectrum should then decrease rapidly 
at low energies. The effect we are searching for might be produced 
by high mass neutralinos with negligible branching ratio into 
$b \bar{b}$ or $t \bar{t}$, which is the case e.g. for a very pure 
heavy Higgsino-like neutralino. 
In Fig.~\ref{fig:speccl1} we show two examples compared to our 
standard background estimate (solid line denoted with the letter $b$).
Model 1 is a 964~GeV Higgsino ($Z_g = 0.0027$) with flux rescaled by 
$f\,\delta = 4180$; it is the model in our MSSM sample which gives
the highest possible flux at 50~GeV compatibly with the background 
normalization we have chosen (dashed line in the figure) and with 
the requirement that added to the former it should give a good fit 
of the {\sc Bess} data.
Model 2 is instead the case in this class of examples which is 
associated with the lowest rescaling, $f\,\delta = 1180$; 
it is at the same time the model with the smallest mass, 
an Higgsino-like neutralino ($Z_g = 10^{-6}$) with 
$m_{\chi} = 777$~GeV. 
A further reduction in the background or a less pronounced 
overproduction of antiprotons at high energies, allow lower values 
of both $m_{\chi}$ and $f\,\delta$.

\begin{figure}[tb]
\resizebox{\hsize}{!}{\includegraphics{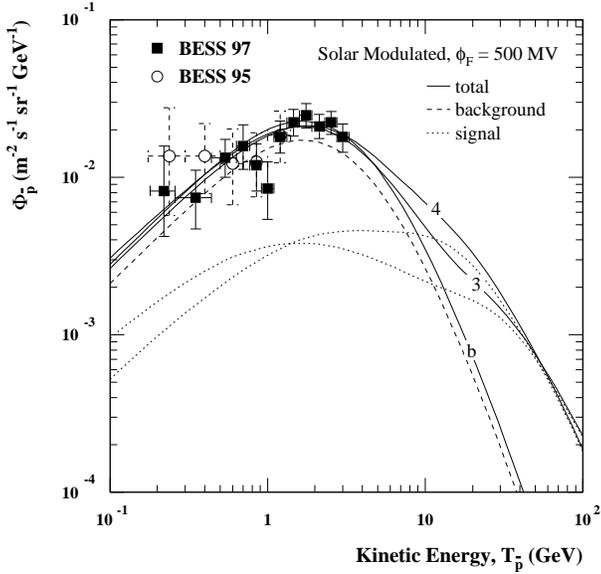}}
\caption{Distorsion on the secondary antiproton flux (line $b$) induced
by a neutralino signal with flat spectrum.}
\label{fig:speccl2}
\end{figure}

\subsection{Break in the energy spectrum}

We focus in this case on rather flat neutralino-induced spectra.
If one adds such a component to the background, the peak and the low 
energy tail in the antiproton flux are not sensibly affected, 
but a severe break may occur in the high energy region. 
A flat signal spectrum is generated overlapping a low contribution 
from a quark annihilation channel to a gauge boson contribution at 
higher energies. 
In Fig.~\ref{fig:speccl2} model 3 is the case in our MSSM sample 
which produces the flattest spectrum and therefore may induce the most 
severe break.
It is a very heavy neutralino, $m_{\chi} = 2730$~GeV, and the 
rescaling needed in this case may be uncomfortably high, 
$f\,\delta \sim 10^5$ (but still it does not violate any 
experimental constraint from other cosmic ray measurements). 
The second example, model 4, has a mass of 1400~GeV and 
$f\,\delta = 7200$.
Clearly these are extreme cases: in this category we may include 
a large fraction of models in our sample, which with a
very mild rescaling give a distorsion of the flux at the highest
energies. The chance of detecting less severe breaks in the energy 
spectrum clearly depends on how accurately the antiproton flux 
is measured.

\subsection{Finite antiproton lifetime}

\begin{figure}[tb]
 \resizebox{\hsize}{!}{\includegraphics{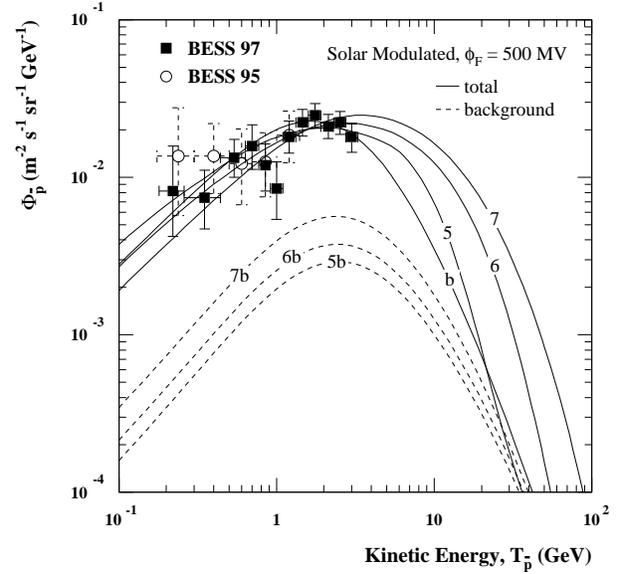}}
\caption{Estimate of possible fluxes in case of finite antiproton
lifetime $\tau_{\bar{p}}$, as compared with the standard background 
case when  $\tau_{\bar{p}} = \infty$. The total flux is the sum of 
a neutralino induced component (not shown) and the secondary 
component (dashed lines) which, as plotted in the 3 examples, 
can be much suppressed compared to line $b$.}
\label{fig:spectau}
\end{figure}

In exotic scenarios, the antiproton may have a finite lifetime 
$\tau_{\bar{p}}$. In the cosmic ray context this is an interesting 
ingredient in case $\tau_{\bar{p}}$ is lower than the characteristic 
escape time of antiprotons from the Galaxy. The escape time is 
determined by the parameters which define the propagation model; 
in our case it is about 10~Myr.
This value is clearly much lower than the bound one can infer in CPT
conserving theories from the experimental limit on the proton lifetime
($\tau_{\bar{p}} = \tau_p > 10^{31}$~yr, see Caso et al. (1998)). 
On the other hand it is much larger than the most stringent direct
experimental bound ($\tau_{\bar{p}} > 0.05$~Myr from the 
$\bar{p} \rightarrow \mu^- \gamma$ decay mode,
see Hu et al. (1998)). 

Geer \& Kennedy (1998) have recently claimed that from cosmic 
ray measurement it is possible to infer a limit of 
$\tau_{\bar{p}} > 0.8$~Myr (90\% C.L.).
In their analysis the secondary component only is included. Adding a 
neutralino signal obviously may change this result.
Actually we show here that in a clumpy scenario, even in case of a 
finite antiproton lifetime, we can produce models which give at the 
same time a very good fit of the existing {\sc Bess} data and 
whose spectral features are peculiar enough to be distinguished 
from the standard background once measurements at higher energies 
will be available. In Fig.~\ref{fig:spectau} we compare three
such models with the case of standard background and infinite 
$\tau_{\bar{p}}$.
Model 5 is a 51~GeV gaugino-like neutralino whose antiproton flux 
has been scaled by $f\,\delta = 49$ and for 
$\tau_{\bar{p}} = 0.82$~Myr. Also shown in the figure is the 
reduction in the background flux induced by considering 
$\tau_{\bar{p}} = 0.82$~Myr (dashed line labelled by $5b$).
Model~7 is the heaviest model for which the predicted flux is still in 
excellent agreement with existing data (90\% C.L. fit for 
$m_{\chi} = 477$~GeV, 
$f\,\delta = 1.2 \cdot 10^4$ and $\tau_{\bar{p}} = 2.92$~Myr), while
model~6 is some intermediate case ($m_{\chi} = 188$~GeV, 
$f\,\delta = 78$ and $\tau_{\bar{p}} = 1.32$~Myr). The trend is that
for heavier neutralinos there is a larger overproduction of 
antiprotons in the high energy range.

Applying the largest possible rescalings consistent with $\gamma$-ray 
measurements we find that a 56~GeV neutralino model gives a flux
which is consistent with {\sc Bess} data at 90\% C.L. in case 
the antiproton lifetime is as low as $\tau_{\bar{p}} = 0.15$~Myr. 
The bound of Geer \& Kennedy (1998) is clearly violated. Notice that 
we are comparing with a more aboundant data set than in that 
reference ({\sc Bess}~97 data were not included there) 
and that we used our standard values for the parameters which define 
the diffusion model and solar modulation. If uncertainties were 
included the lower bound we would get with this method would 
probably be very close to the most stringent direct experimental bound 
$\tau_{\bar{p}} > 0.05$~Myr or lower.

\section{Conclusion}

To conclude, we have shown that there is a chance of detecting
neutralino dark matter in upcoming measurements of the cosmic
antiproton flux at high energies. The signatures we propose here
are alternative to the signature of an exotic component at low
kinetic energies, which seems not to be required by present data.
We have also discussed the possibility that antiprotons have a finite
lifetime and shown that the limit on $\tau_{\bar{p}}$ 
which is possible to set 
on the basis of cosmic ray measurements
is comparable to those in direct experiments.

\bigskip

\noindent{\bf Acknowlegements}

\smallskip 

\noindent
I am grateful to Lars Bergstr{\"o}m and Joakim Edsj{\"o} for many
useful discussions. I thank Paolo Gondolo for collaboration on the
numerical supersymmetry calculations.

\end{document}